\documentclass[a4paper]{article}

\usepackage{multicol,amsmath,amssymb,graphicx,float}
\usepackage[a4paper,left=25mm,right=25mm,top=25mm,bottom=30mm]{geometry}
\usepackage[auth-lg,affil-it]{authblk}
\usepackage[hidelinks]{hyperref}
\usepackage{cite}
\usepackage{titlesec}

\title{Morphogenesis of filaments growing in flexible confinements}

\author{Roman Vetter}
\author{Falk K.~Wittel}
\author{Hans J.~Herrmann}
\affil{Computational Physics for Engineering Materials, IfB, ETH Zurich, Stefano-Franscini-Platz 3, CH-8093 Zurich, Switzerland}

\begin{document}

\maketitle

\begin{abstract}
Space-saving design is a requirement that is encountered in biological systems and the development of modern technological devices alike. Many living organisms dynamically pack their polymer chains, filaments or membranes inside of deformable vesicles or soft tissue like cell walls, chorions, and buds. Surprisingly little is known about morphogenesis due to growth in flexible confinements---perhaps owing to the daunting complexity lying in the nonlinear feedback between packed material and expandable cavity. Here we show by experiments and simulations how geometric and material properties lead to a plethora of morphologies when elastic filaments are growing far beyond the equilibrium size of a flexible thin sheet they are confined in. Depending on friction, sheet flexibility and thickness, we identify four distinct morphological phases emerging from bifurcation and present the corresponding phase diagram. Four order parameters quantifying the transitions between these phases are proposed.
\end{abstract}

\begin{multicols}{2}

\subsection*{Introduction}

What morphologies will a thin object adopt when subjected to tight spatial confinement? This fundamental question of morphogenesis arises on a large range of length scales, from DNA strands packed in bacteriophage capsids and globules \cite{STSGAB01,KTBG01,O04} to the folding of pop-up tents \cite{MCPRJ12}. Packing problems have played a major role in the understanding of spatial self-organization in living organisms and technological applications alike. Thin sheets of foil or paper, for instance, develop complex ridge networks when folded and crumpled \cite{LGLMW95,BAP97,BK05,VG06,TAT08,TAT09}.

Significant progress has recently been made in the understanding of dense packings of elastic and elasto-plastic wires, in absence of thermal fluctuations, inside of rigid three-dimensional confinement \cite{GBA08,SNWHH11,NSWH12,VWSH13}. A particular restriction shared by all these studies is the perfect rigidity of the cavities---a constraint rarely met in nature or biomedical applications. Detachable platinum coils for example, which have revolutionized the surgical treatment of saccular cerebral aneurysms \cite{GVSM91}, are many orders of magnitude stiffer than the arterial walls they are fed into \cite{SAKR89}. Microtubules confined in lipid bilayer membranes \cite{EKFL96,KFML97} and erythrocytes \cite{NJSC83,CSS98} as well as actin/filamin networks in vesicles \cite{HTIH99,LS02} are able to deform their weak confinements significantly. In turn, such cavities force the contained filaments to buckle and reorder if their persistence length grows large enough. Recent experiments on coiled elastic nanowires and nanotubes encapsulated in swelling polymer shells and emulsion droplets have demonstrated how mechanical work can be stored and deployed through deformable spatial confinement \cite{XWLYCWWLXC10,CWXSYZZZC11,CYWXLCC13}.

Knowing the mechanisms that govern morphogenesis in constrained spaces is key to understanding the ordering in dense packing problems. Here, we present computer simulations and table-top experiments with everyday materials carried out to explore the morphological phase space of filament packings in a regime that is conceptually very different from previous systematic studies. A regime where the growing filament can strongly deform the cavity, allowing for a surprisingly enriching mutual feedback between the two structures. A decisive set of two geometric and two material parameters are identified that control this complex interaction, giving rise to no less than four distinct morphologies vastly differing in their packing evolution and energetics. Our findings provide the foundation for a new research field in which packing processes and growth are governed by the complex interplay of thin deformable bodies made of different materials. The reported observations exemplify the dramatic influence of friction and confinement rigidity on packing processes in both natural and technological applications.

\subsection*{Model and parametrization}

\begin{figure}[H]
\includegraphics[width=\columnwidth]{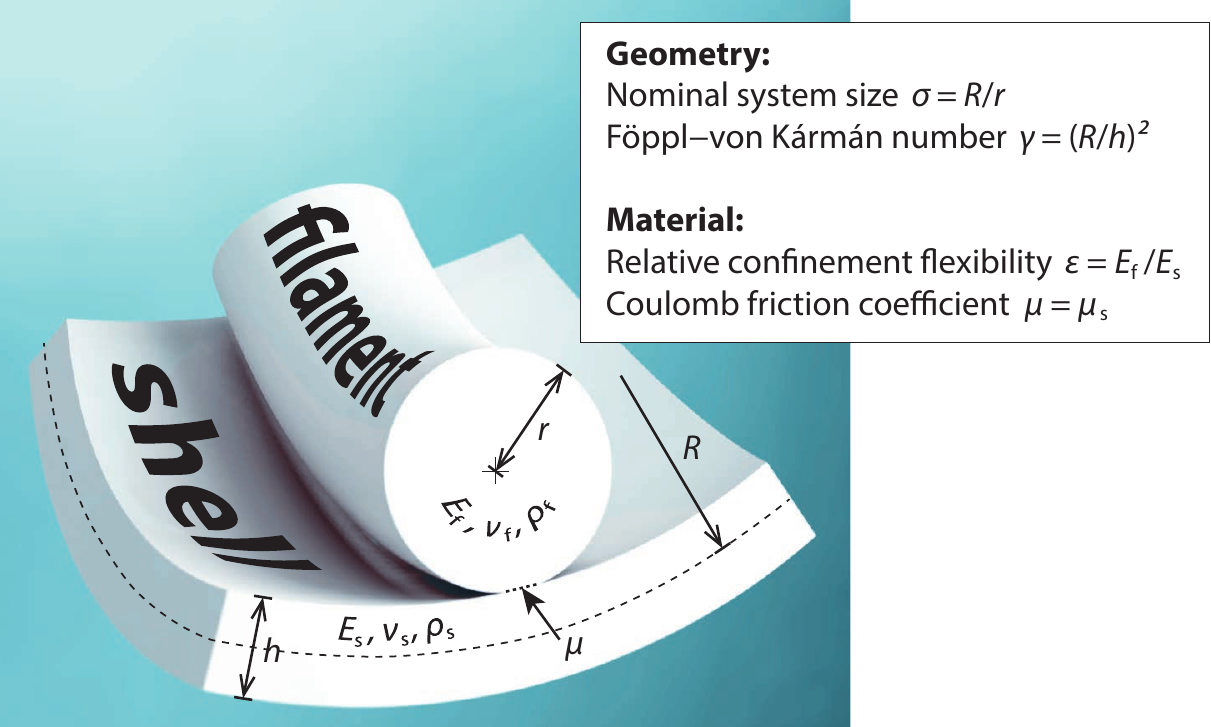}
	\caption{\textbf{Schematic of the interaction model.} Both the circular filament and the confining spherical sheet are characterized by a Young's modulus $E$, Poisson's ratio $\nu$ and mass density $\rho$. We assume stick-slip Coulomb friction between any contacting surfaces. The morphogenesis is governed by only four dimensionless numbers: $\sigma$, $\gamma$, $\varepsilon$, $\mu$.}
	\label{fig:sketch}
\end{figure}

The most intelligible and pure way of gaining quantitative insight into morphogenesis in flexible confinements is by continuum-mechanical simulations with linearly elastic materials. Many of the aforementioned applications are best matched by a circular ring filament growing inside of a closed spherical thin sheet. Figure \ref{fig:sketch} shows a cutout of our computational model (see also \hyperref[sec:suppl1]{Supplementary Note 1}). Both filament and sheet are characterized by three homogeneous isotropic material parameters, labelled with subscripts f and s, respectively: the Young's modulus $E$, Poisson's ratio $\nu$, and mass density $\rho$. In accordance with a suggestion for condensed DNA \cite{O04}, Coulomb's law is assumed for dry stick-slip friction between any two contacting surfaces (filament-filament, filament-shell, shell-shell) with an isotropic static friction coefficient, $\mu_\mathrm{s}$, and a slightly lower dynamic friction coefficient, $\mu_\mathrm{d}$. To simplify the parameter space, we use the same coefficients for all three contact types and fix $\mu_\mathrm{d}=0.9\mu_\mathrm{s}$, which is adequate for a broad class of materials \cite{FLS64}.

\begin{figure*}[ht]
	\centering
	\includegraphics[width=\textwidth]{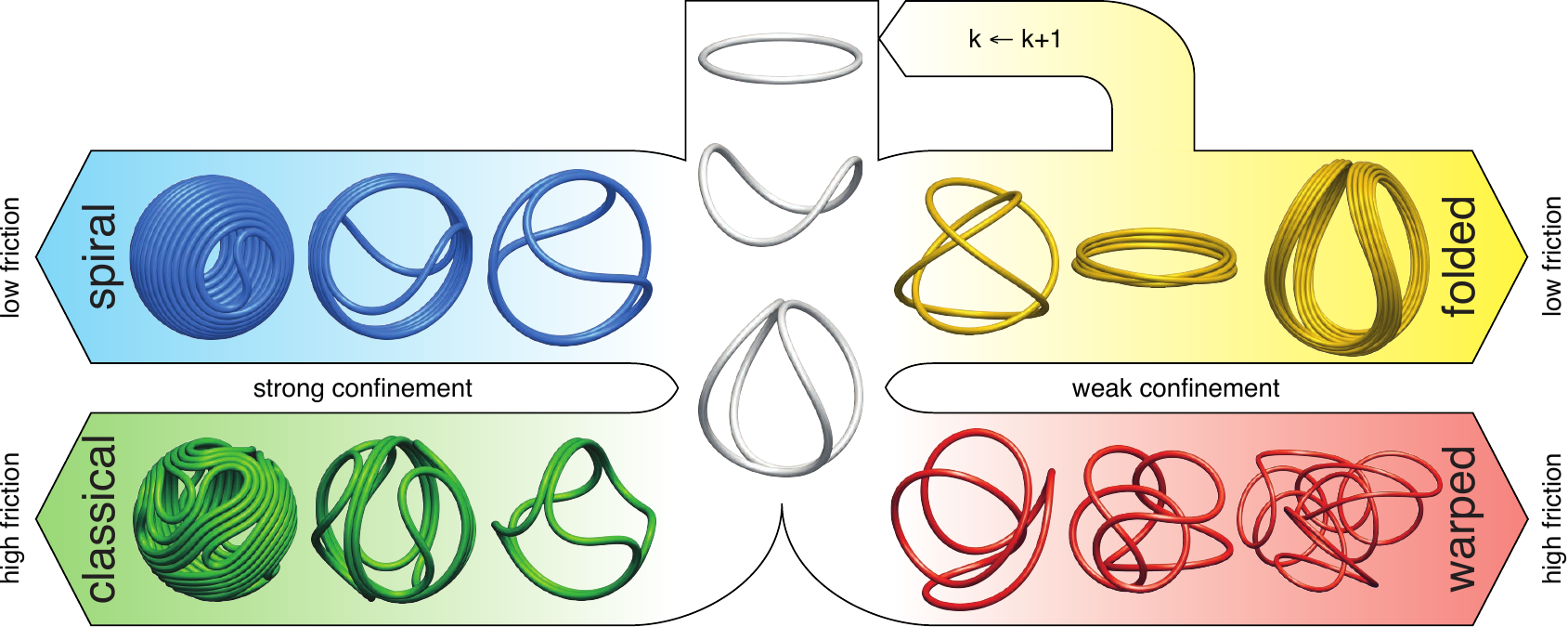}
	\caption{\textbf{Packing evolution depending on friction and confinement rigidity.} A confined growing ring buckles to a saddle shape (buckling mode $m=2$, middle). Beyond the point of first contact $l=l^{*}$, four distinct morphologies can emerge. The folding at low friction and weak confinement is repeated during growth: Filament bundles refold self-similary, each time tripling the number of bundle strands ($n=3^k$).}
	\label{fig:phases}
\end{figure*}

The initial condition at time $t=0$ consists of a ring filament with length $L$ and cross-sectional radius $r$, surrounded by a close-fitting spherical shell with thickness $h$, whose middle surface has radius $R$. As the filament grows, it bears against the confining wall until the critical buckling load $4E_{\mathrm f}I_{\mathrm f}/R_{\mathrm{b}}^2$ (where $I_{\mathrm f}=\pi r^4/4$ is the second moment of area) is exceeded and it buckles out of plane with harmonic mode $m$. Excited modes $m>2$ are unstable in the quasi-static frictionless limit \cite{GV12}, but our simulations show that they can be observed when inertia is not negligible  (i.e., large mass densities, fast growth or viscous overdamping) and in the presence of significant friction. The ground state buckling mode $m=2$, which is most relevant in practice, develops a saddle shape as depicted in the middle of Fig.~\ref{fig:phases} until two filament segment pairs touch. This first contact occurs at $l := L(t)/L(0) = l^{*}\approx 2.127$ for rigid spherical cavities in the theoretical thin filament limit $r\to 0$ \cite{GV12}. The bending energy $U_{\mathrm b}$ of the filament before $l=l^{*}$ can be approximated analytically by
\begin{equation}
\frac{U_{\mathrm b}R_{\mathrm f}}{E_{\mathrm f}I_{\mathrm f}}=\frac{1}{2}\int_0^{2\pi l}\mathrm ds\,\left[\kappa(s)^2+1\right],
\label{eq:ebend_theo1}
\end{equation}
where $\kappa$ is the geodesic curvature of the rim of a unit \textit{excess cone} (e-cone) \cite{MBG08}, and $R_{\mathrm f}:=L(0)/2\pi=R-h/2-r$ is the effective filament radius. Growth beyond $l^{*}$ in finite systems with real self-avoiding materials, however, has never been explored to date. Our simulations and experiments dispel this limitation, showing that four distinct morphologies emerge by bifurcation as the filament grows longer (Fig.~\ref{fig:phases}). We denominate them \textit{spiral}, \textit{classical}, \textit{folded} and \textit{warped}, motivated by their characteristics as detailed in the following.

A key result from the computer simulations is that this morphogenesis is controlled by four independent dimensionless non-negative quantities,
\begin{equation}
\sigma=R/r,\quad \gamma=(R/h)^2, \quad \varepsilon=E_{\mathrm f}/E_{\mathrm s},\quad \mu=\mu_{\mathrm s},
\end{equation}
if inertial effects are negligible. Up to an irrelevant prefactor, $\gamma$ is the F\"oppl--von K\'arm\'an number, a geometrical measure for a thin sheet's tendency to bend rather than stretch. $\varepsilon$ is the relative filament rigidity, which conversely may also be thought of as the confinement flexibility.  $\varepsilon$ and $\gamma$ generalize the previously studied rigid cavities, which are attained in the limits $\varepsilon\to 0$ and $\gamma\to 0$. Together with $\gamma$, expressing the nominal system size by the non-dimensional ratio $\sigma$ renders the problem scale-invariant, which attests to the wide applicability of our results from microscopic to macroscopic scales.

\subsection*{Morphological phases and phase transitions}

\begin{figure*}
\includegraphics[width=\textwidth]{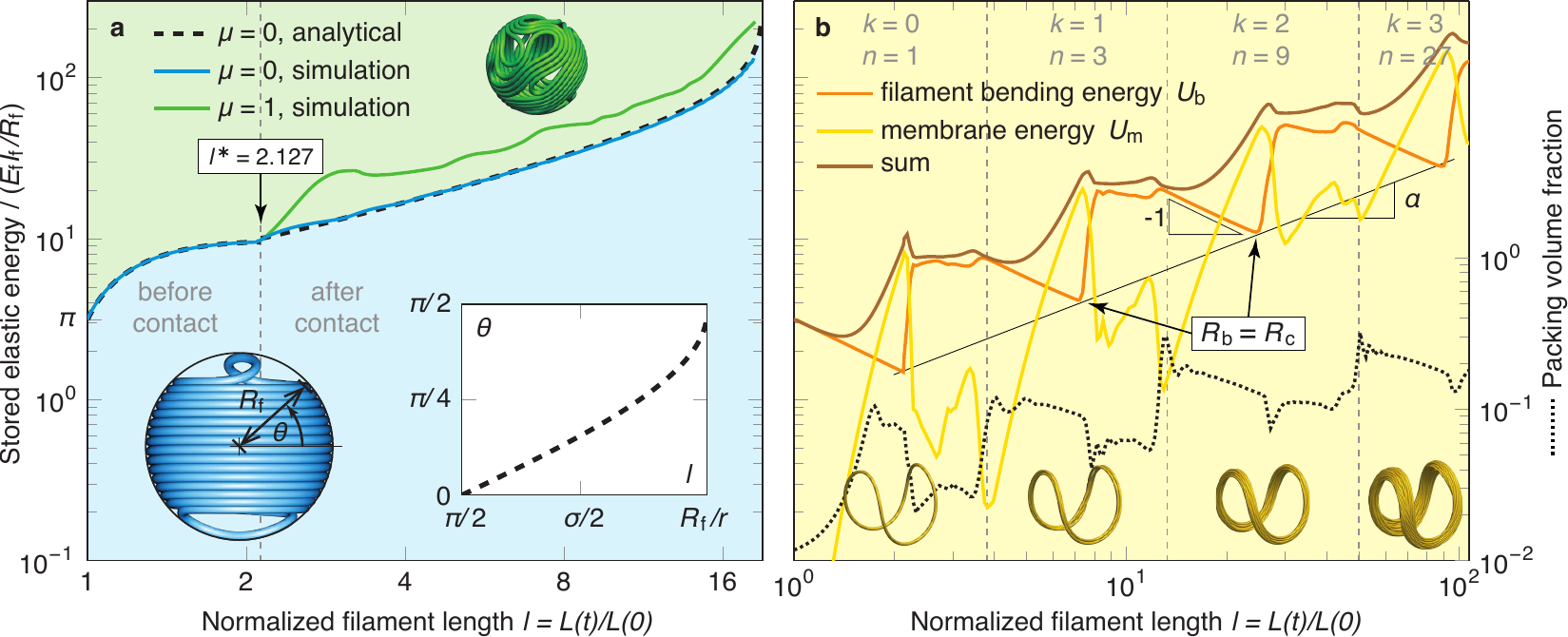}
	\caption{\textbf{Comparison of dynamics and energetics.} {\bf a}, Filament bending energy in the spiral (blue) and classical (green) morphology at $\sigma=20$. The inset shows the inclination $\theta(l)$ of the spiral coil. {\bf b}, Folded morphology at $\sigma=20$, $\gamma=20^2$, $\varepsilon=10^4$, $\mu=0$. The predominant elastic energy contribution alternates between the filament bending energy $U_{\mathrm{b}}$ and the membrane energy $U_{\mathrm m}$ (stretching term of equation (\ref{eq:shell})). A cascade of four self-similar folds can be recognized by the repeated power law regions.}
	\label{fig:energy}
\end{figure*}

The packing in frictionless rigid spheres can serve as a toy model for less idealized systems. When $l>l^{*}$, a spiral (depicted in Fig.~\ref{fig:phases}, top left) develops analogous to unconfined e-cones \cite{SWBMH10}. In the spirit of refs~\cite{BABCCET06,SWH08}, we call this the \textit{spiral phase}. Simple geometrical arguments (see \hyperref[sec:suppl2]{Supplementary Note 2}) allow for an analytical approximation of the predominating elastic contribution, the bending energy $U_{\mathrm{b}}$ of the confined filament:
\begin{equation}
\frac{U_{\mathrm b}R_{\mathrm f}}{E_{\mathrm f}I_{\mathrm f}}\approx\pi\frac{R_{\mathrm f}}{r}\log\left[1+\frac{2}{\cot(\theta/2)-1}\right]+\frac{4\pi}{\pi/2-\theta},
\label{eq:ebend_theo2}
\end{equation}
where the coil inclination $\theta$ is implicitly determined by $l\approx\sin(\theta)R_{\mathrm f}/r+\pi/2-\theta$. In Fig.~\ref{fig:energy}a we show that equations (\ref{eq:ebend_theo1}) and (\ref{eq:ebend_theo2}) are in excellent agreement with our numerical measurements. The measured energy weakly oscillates because the coil is slightly bent by the S-curves. These oscillations increase for larger $\sigma$. As the surface gets fully covered with a single layer of filament ($l$ approaching $R_{\mathrm f}/r$), the growing filament eventually buckles inward to release a large amount of elastic energy, and the packing process continues in a less ordered fashion much like some DNA molecules in phage capsids \cite{KTBG01,ATVSH02}.

\begin{figure}[H]
\includegraphics[width=\columnwidth]{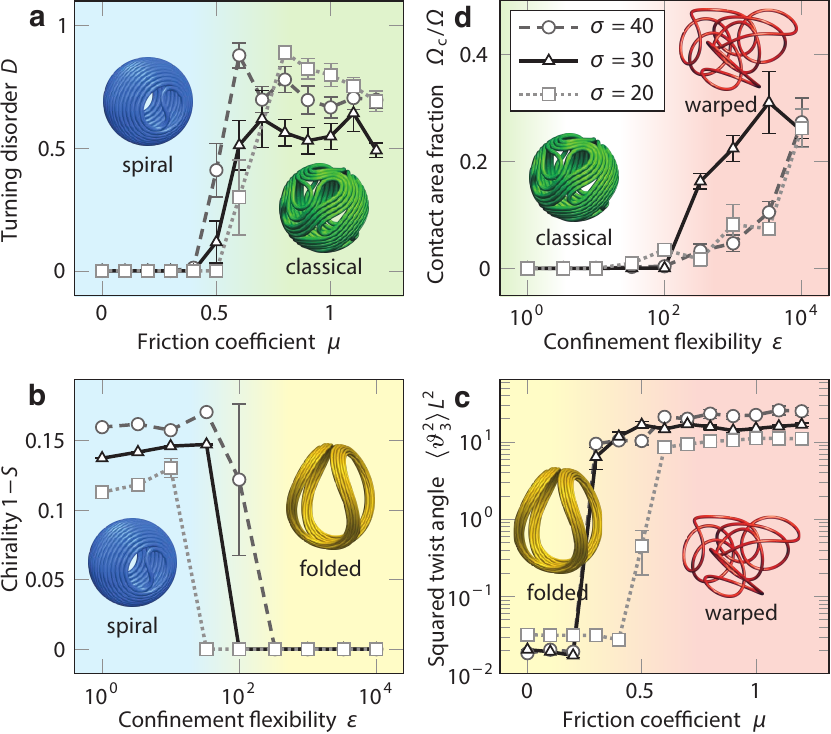}
	\caption{\textbf{Order parameters for the transition between the phases.} {\bf a}, Half of the spiral filament turns right and half of it left, which is not the case in the classical phase ($\varepsilon=0$, $l=10$). {\bf b}, The dissimilarity between the filament and its own mirror image is strictly positive in the chiral spiral phase, but vanishes in the achiral folded phase ($\gamma=10^4$, $\mu=0$, $l=5$). {\bf c}, The amount of torsion quantifies the transition from the folded to the warped phase ($\gamma=\varepsilon=10^4$, $l=5$). {\bf d}, The fraction of the sheet surface in contact with itself vanishes when the warped phase transitions to the classical phase ($\gamma=10^4$, $\mu=1.4$, data maximized over $l\in[1,10]$). Error bars represent standard errors from 6-10 independent realizations.}
	\label{fig:order_parameters}
\end{figure}

A crucial requirement for a growing filament to coil is tangential sliding, giving rise to the high degree of order in the spiral morphology by continuous rearrangements. Many practical materials however resist sliding by frictional forces and even DNA does so \cite{KCDSD05,G12,OSLKK13}. Friction induces locality by limiting rearrangments to the local neighbourhood, inhibiting relaxations to lower global energy conformations, thus trapping growing filaments in a disordered state (Fig.~\ref{fig:phases}, bottom left). Figure \ref{fig:energy}a clearly shows that the bending energy is bounded from below by the spiral phase. A similar situation is encountered in flat wire packings \cite{SWH08}, from where we adopt the term \textit{classical phase}. To quantify the spontaneous breaking of spiral symmetry with increasing $\mu$, we propose the following order parameter. Let $[p_1, p_3, p_3]^{\mathrm T}(s)$ denote the position of the centreline of the filament after transformation onto its own principal axes, in order of descending principal moments of inertia. Then, the sign of $\tau(s)=\partial_s^2p_2\partial_sp_1-\partial_s^2p_1\partial_sp_2$ indicates its turning direction as seen along the axis of minimal moment of inertia. Consequently, the circular convolution
\begin{equation}
c(q)=\frac{1}{L}\int_0^L\mathrm ds\;\mathrm{sgn}\big[(s-L/2)\,\tau(s+q)\big]\in[0,1]
\end{equation}
is a natural measure for the degree of order of the filament when cut in half at $s=q$ and $s=q+L/2$. Here, we implicitly extend $\tau(s)=\tau(s+L)$ periodically. If the two halves equally contain right- and left-turning segments, we have $c(q)=0$. Conversely, only if one half turns only left and the other half only right, we have $c(q)=1$. Thus, if we account for the periodic nature of the ring filament by maximizing over all bisection points $q$, the turning disorder $D:=1-\max\{c(q)|q\in[0,L]\}$ can serve as an order parameter to discriminate the spiral from the classical phase. It is evident from Fig.~\ref{fig:order_parameters}a that the transition occurs near $\mu\approx 0.5$, with a slight dependence on $\sigma$. In stiff confinement, friction must thus be fairly strong to introduce local order, which contributes to explaining why viral DNA is often condensed into layered spools \cite{EH77,CCRMBS97,OGEB01,JCJWKC06}.

The surface-covering spiral and classical morphologies bear resemblance to liquid crystals (LCs), an analogy that has already been drawn in the context of DNA packing in viral capsids \cite{KAB06}. Let's define a loop by an area surrounded by a filament segment with only one inner point of contact \cite{SWH08}. The spiral phase has only four such loops (two at each pole), while the classical phase is characterized by a broader spatial distribution of loops. At very high surface packing ($l\to R_{\mathrm f}/r$), classical loops are compressed to point singularities with strength $1/2$ delimiting line disclinations known from nematic LCs \cite{C92}. The spiral phase is in turn reminiscent of spherical smectic LCs with two closely bound disclinations ending at two 1/2-singularities at each pole \cite{JCLXAXBL09}. The total disclination strength of a spherical LC is always two, which is a direct consequence of the Gauss--Bonnet theorem. Indeed, this identity holds for growing ring filaments for all $\sigma,\mu$ in the spiral and classical phases. The number of topologically nontrivial loops is always four, but the classical phase can exhibit an arbitrary additional number of topologically trivial loops. However, we stress that the existence of topologically trivial loops is sufficient, but not necessary for $D>0$, and is thus not an order parameter.

The spiral and classical morphologies are highly metastable as the filament inevitably buckles away from the rigid wall. A qualitative stability condition was derived in ref.~\cite{KAB06} and translates to $l\ll \sigma^2$ in our terms, suggesting that dense single-layered surface packings are found only in sufficiently large cavities. This explains why they are not common in biophysical environments and vesicles. Instead, such systems gain ultimate stability from weak or flexible confinement. In the computer simulations, we increased $\gamma$ and $\varepsilon$ to discover a completely altered morphogenesis beyond a certain transition. As the confining sheet is weakened or thinned, suddenly, the filament folds on itself as illustrated in the top right of Fig.~\ref{fig:phases} instead of coiling. Bundles of $n$ subthreads are formed similar to actin/filamin rings and microtubules in vesicles \cite{HTIH99,LS02,PCKANG09} or folded poles in pop-up tents \cite{MCPRJ12}. We hence refer to this as the \textit{folded phase}. The folding process is repeated as the filament continues to grow. The winding number $n$ obeys $n=\prod_k(2m_k-1)$, where $m_k\in\{2,3,4,...\}$ is the buckling mode and $k=0,1,2,...$ is the number of folds in the cascade (see also \hyperref[sec:suppl3]{Supplementary Note 3}). The ground state ($m_k\equiv 2$) energy is plotted in Fig.~\ref{fig:energy}b, revealing a series of self-similar folds that define the folded phase. Prior to buckling, the circular filament bundle with bulk radius $R_{\mathrm{b}}$ expands to release bending energy according to $U_{\mathrm{b}}R_{\mathrm f}/E_{\mathrm f}I_{\mathrm f}\approx n\pi R_{\mathrm f}/R_{\mathrm{b}}\sim l^{-1}$, stretching the circumjacent sheet until a critical radius $R_{\mathrm{b}}=R_{\mathrm{c}}$ is reached where the bundle buckles. The long-term trend is $U_{\mathrm{b}}R_{\mathrm f}/E_{\mathrm f}I_{\mathrm f}\sim l^{\alpha}$ with $\alpha\in(0,1)$, indicating that the critical radius increases as $R_{\mathrm{c}}/R_{\mathrm f}\sim l^{(1-\alpha)/2}$. For the set of parameters in Fig.~\ref{fig:energy}b, $\alpha=0.80\pm0.02$. This striking refolding of bundled rings provides a purely mechanical explanation for the spontaneous bundling of weakly confined filaments such as actin networks \cite{HTIH99,LS02} and marginal microtubule bands in developing erythrocytes \cite{NJSC83,CSS98} as a result of membrane or shell enclosure instead of cross-linkage. Only sufficiently deformable containers conform to pushing filaments, allowing them to pass one another to fold into energetically more favourable bundle configurations. This might also provide a paradigm to explain layered slime thread bundling in hagfish gland thread cells \cite{WHMLDBBHVTF14}, where the cell membrane deforms under high packing pressure. Our findings suggest that such thread bundle packings may be obtained only in systems where frictional forces are rather small. Other biological systems in which filament bundling provides a mechanism of mechanical stabilization in membrane confinement include filopodial protrusion \cite{SBCVKVB03,MR05}.

Remarkably, the gain in mechanical stability in flexible cavities goes hand in hand with the loss of chirality. A convenient order parameter describing the transition from the spiral to the folded phase is the chirality measure $1-S$, where $S$ is the degree of similarity of the filament with its own mirror image as defined in ref.~\cite{MR91}, maximized over all possible mirror planes. In Fig.~\ref{fig:order_parameters}b, a pronounced discontinuity in $1-S$ is in evidence, suggesting that the phase transition is of first order. Chirality of confined rods is know to play a key role in morphogenesis of \textit{Escherichia coli} cells for instance, where the bacterial rod grows into a helical spiral, guided by proteins \cite{WFHS12}. Our results provide evidence that, conversely, filament chirality can emerge as a purely mechanical consequence of non-flexible confinement.

If growing filaments are inclined to form highly ordered bundles inside of deformable membrane cavities (without friction), the question naturally arises whether purely mechanical material properties can also give rise to disordered packing morphogenesis with strongly warped or tangled filaments. Developing vertebrate intestines, where the gut tube grows into the body cavity at a different rate than the adhering mesenteric sheet, are in fact one example \cite{SKSFLMT11}. Here, we are able to put this particular result in a broader, more general framework by controlling friction. As $\mu$ is increased, the thin flexible sheet grabs hold of the pushing filament, tightly wrapping around it. Just like in the classical phase, friction (or adhesion) enforces locality: The filament can no longer just freely fold up inside, and further growth causes it to locally twist in frustration, leading to a \textit{warped} morphology (depicted in Fig.~\ref{fig:phases}, bottom right). For $\gamma,\mu\to\infty$, this behaviour is reminiscent of the Euler--Plateau problem \cite{GM12} but significantly more complex due to the crucial role of twist and volumetric exclusion. The transition from the folded to the warped phase is accompanied by the breaking of torsional symmetry. As order parameter the non-dimensional torsional energy $\langle\vartheta_3^2\rangle L^2=2LU_{\mathrm t}/G_{\mathrm f}J_{\mathrm f}$ (see \hyperref[sec:suppl1]{Supplementary Note 1}) may be used, since it vanishes in the folded phase for $\sigma\to\infty$ and takes a significant, finite value in the warped phase. As can be recognized from Fig.~\ref{fig:order_parameters}c, the exact value of the corresponding critical friction coefficient, while depending on the system size $\sigma$, is generally very low, implying that the warped phase is relevant even in systems with moderate friction. Within the warped phase, the stored torsional energy quickly levels off. Our measurements thus provide a tight lower bound for the amount of twist in various vetrebrate guts \cite{SKSFLMT11}.

The direct transition from the classical to the warped morphology is less obvious. A large portion of the phase space is occupied by \textit{mixed} states in which the confinement is not stiff enough to keep the filament from buckling into the sphere, but at the same time not flexible enough to wrap around it and force it to twist. Such configurations are prevalent, e.g., in brain aneurysms occluded by detachable coils (see e.g.~ref.~\cite{RS07}). It is nonetheless possible to define a sharp phase boundary (PB) by considering as an order parameter the area fraction of the sheet that is in contact with itself, $\Omega_{\mathrm c}/\Omega$, because it is non-zero only in the warped phase (Fig. \ref{fig:order_parameters}d).

\subsection*{Morphological phase diagram}

\begin{figure*}[ht]
	\includegraphics[width=\textwidth]{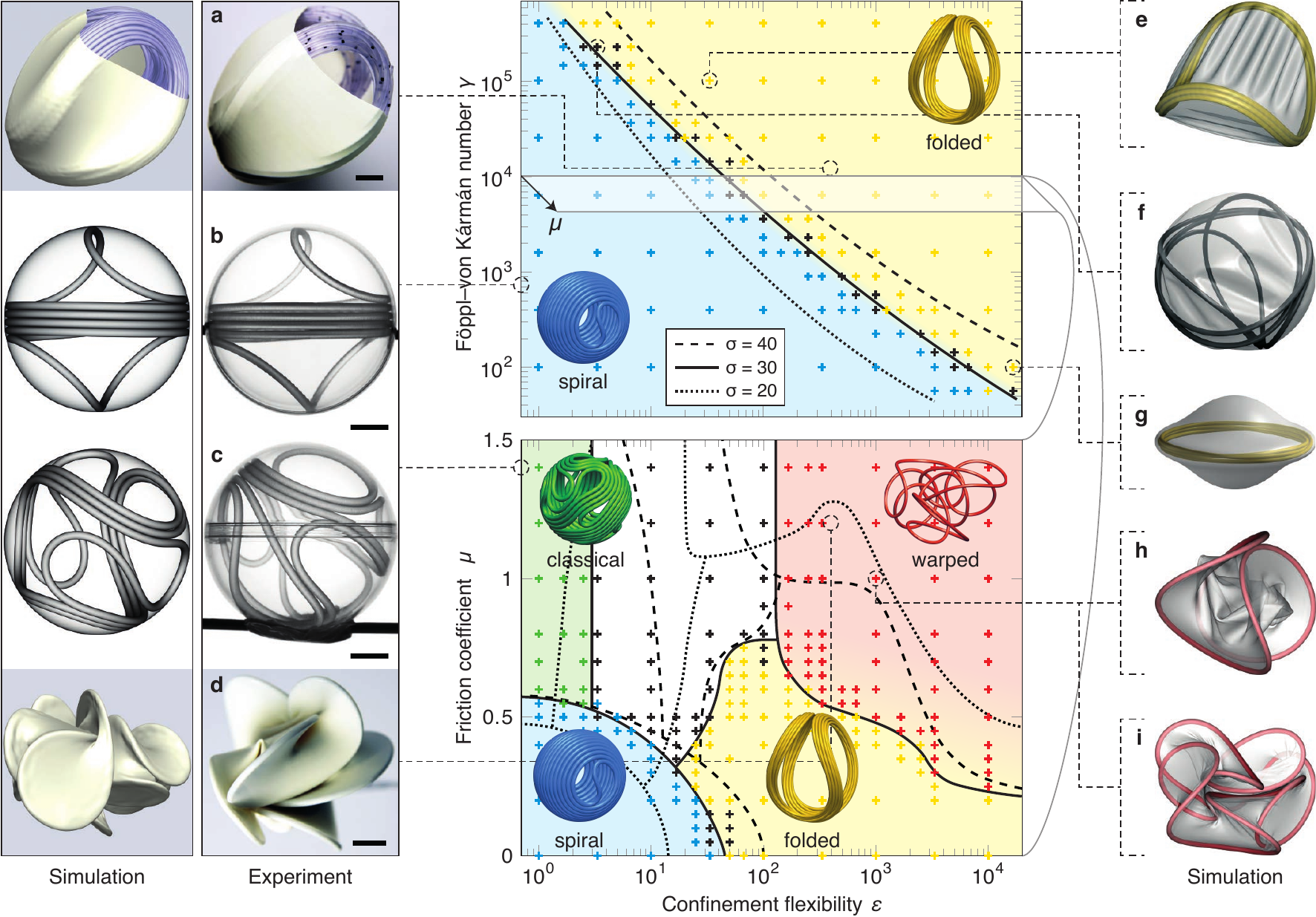}
	\caption{\textbf{Morphological phase diagram.} For $\sigma=30$, individual coloured symbols denote numerical simulations producing the different phases and black symbols are mixed (see text). The lines represent least-squares-fitted phase boundaries. The fits for $\sigma=20,40$ were obtained with equivalent parameter resolution (individual realizations not shown). {\bf b},{\bf c}, Polyurethane wires ($r=1$ mm) tangentially fed from opposite directions into rigid polystyrene spheres ($R=24$ mm, $h=1$ mm). {\bf a},{\bf d}, Polycaprolactam wires ($r=0.75$ mm) in natural rubber balloons ($R=27$ mm, $h=0.25$ mm). {\bf e}-{\bf i}, Simulation snapshots at $\sigma=30$, $l=4$. Very thin confinements exhibit tension wrinkles \cite{CRM02} ({\bf e}). {\bf f} shows an exemplary mixed configuration close to the phase boundary. Very flexible confinements are stretched similar to lipid vesicles \cite{KFML97} ({\bf g}). Warped filaments typically first crumple the sheet ({\bf h}, $l=2$) before strongly twisting ({\bf i}, $l=4$). The scale bars are 1 cm.}
	\label{fig:phase_diagram}
\end{figure*}

For the design and optimization of new materials and structures, e.g.~in nanorobotics or endovascular coiling, reliably predicting the packing behaviour is essential. We used our computer simulations to acquire a quantitative image of the morphological phase space, allowing us to accurately predict morphogenesis depending on the geometry and material parameters. In the low friction regime (upper half of Fig.~\ref{fig:phase_diagram}), the spiral and folded phases are separated by a smooth PB which is well approximated by quadratic curves in $\log(\varepsilon)$-$\log(\gamma)$ space. Evidently, if the confining sheet is thin enough, filaments don't need to be substantially more rigid to fold instead of coiling. Bigger systems (larger $\sigma$) favour the spiral morphology as the PB is shifted toward weak confinements (large $\gamma$, $\varepsilon$). Very close to the PB, mixed configurations such as the one shown in Fig.~\ref{fig:phase_diagram}f occur when the filament folds at only one of the two contact points, which is a dynamic effect.

A cut through the phase space at fixed $\sigma$ and $\gamma$ unveils its full complexity (lower half of Fig.~\ref{fig:phase_diagram}). All phase boundaries are $\sigma$-dependent, and the spiral and classical morphologies prevail toward large $\sigma$ as predicted by the stability condition. Perhaps most intriguingly, we find straight single-parameter lines along which all morphologies are traversed, including the mixed region (uncoloured area). An example is the line along the $\varepsilon$-axis, at $\sigma=40$, $\gamma=10^4$, $\mu=0.5$. This shows how delicate the choice of parameters is for targeting a specific morphology---possibly too delicate for nature to rely on this selection in some parameter regions. Another striking feature is reentrancy of the warped phase in small systems with strong friction (Fig.~\ref{fig:phase_diagram} at $\sigma=20$, $\gamma=10^4$, $\varepsilon\approx 10^2-10^3$, $\mu\approx 1.2$), where the folded phase extends far into the large-$\mu$ region. Anyway, in its low-$\mu$ end near the boundary to the folded phase, the warped morphology is just a temporary interstate. The filament first warps (including the characteristic twist and sheet-sheet contact), but upon further growth, some sliding allows it to rearrange and fold nevertheless, defining a region in phase space where the warped and folded phases coexist, which is illustrated by a colour gradient in the phase diagram.

\subsection*{Discussion}

Natural shell tissue is often softer than other thin bodies it gets in contact with. We have emphasized the importance of taking this flexibility into account in the study of morphogenesis in spatially confined systems. Depending on the involved size ratios and friction, the confinement need not be substantially less rigid to allow for a very rich nonlinear mutual interaction, giving way to completely changed morphogenesis. We have identified four morphological phases in the linear growth of confined thin filaments, building a bridge between thepacking of confined DNA, the growth of guts, the bundling of actin networks in cell-sized liposomes, and even liquid crystals. Our results from computer simulations are fully consistent with experiments we conducted on off-the-shelf materials at the human length scale (Fig.~\ref{fig:phase_diagram}a-d).

The presented morphological phase diagram is independent of how growth is realized in detail. In the simulations we grew the filament uniformly everywhere, whereas the tangential injection of an invariant wire in the experiments corresponds to concentrated growth at the point of insertion. These two extremes are exactly equivalent in the low friction phases owing to global rearrangements, and they similarly produce the high-friction morphologies with the exception that reorganization is somewhat condensed to a neighbourhood about the growth/injection zone. Even the converse problem, an invariant filament getting gradually compressed by a shrinking shell, yields the same morphologies as simulations revealed. The only difference is that $\sigma$ (and possibly $\gamma$) decrease over time and the phase diagram must be interpreted accordingly. All presented analytical and scaling arguments hold also for this case without modification (where $R_{\mathrm f}$ is no longer constant, and $l:=L/2\pi R_{\mathrm f}$).

Our findings establish a paradigm for understanding morphogenesis of thin filaments in a multitude of biological mechanisms. Most importantly, we showed how nature may employ flexible envelopment and low frictional forces as a mechanical trick to realize spontaneous bundling and alignment of confined threads, as it is observed in giant vesicles \cite{HTIH99,LS02}, erythrocytes \cite{NJSC83,CSS98}, hagfish cells \cite{WHMLDBBHVTF14} etc., without need for filament interlinking. On the technological side, the morphologies we discovered in flexible confinement should find direct impact in nanorobotics and nanomotors, for which the reported folding of elastic rings provides a new method to stably store and deploy mechanical work in tightly confined spaces. Unlike open nanowires, which coil into quasi-two-dimensional spirals \cite{XWLYCWWLXC10}, ring-like filaments fold in three dimensions and possess no sharp ends that could pierce their environment. Such systems need to be designed with as little friction as possible in order to avoid the energetically and spatially less optimal warped phase.

\footnotesize

\subsection*{Computer simulations}

We minimized the total elastic energy numerically with the finite element method. The filament was modelled by locking-free beam elements with an exact shear representation \cite{RWL97} embedded into a co-rotational formulation \cite{C90} for geometric nonlinearity, while the shell was represented by Loop subdivision surface elements \cite{COS00} that provide the $C^1$-continuity required for boundedness of the curvature integral in equation (\ref{eq:shell}). These descriptions have been published in full detail elsewhere \cite{VWSH13,VSJWH13}. Up to 4000 elements were used to discretize the filament, whereas the shell consisted of 20480 triangles. We let the filament grow uniformly in length according to $l:=L(t)/L(0)=\exp(\lambda t)$. The growth rate $\lambda$ was set sufficiently small for inertial effects to have a negligible effect on the outcome. Volumetric exclusion was imposed by repulsive normal forces upon overlap according to the Kelvin--Voigt model where the elastic part was determined using Hertzian contact theory. A dry slip-stick friction model \cite{M05} was applied for the tangential Coulomb forces. Inertial terms with mass densities $\rho_{\mathrm f}$ and $\rho_{\mathrm s}$ to account for dynamics as well as small subcritical viscous damping for equilibration were added. Newton's equations of motion were integrated in time with the predictor-corrector constant-average acceleration method and adaptive time-stepping. A tiny random perturbative deflection was imposed on the ring filament to allow it to break the system's initial symmetry by Euler buckling and to allow for independent repetitions of the simulations. Our aim was to simulate realistic yet generic materials, and thus we chose $2r=1$ mm, $E_{\mathrm f}$ = 1 GPa, $\rho_{\mathrm f} = \rho_{\mathrm s}$ = 1 g cm$^{-3}$, $\nu_{\mathrm f}=\nu_{\mathrm s}=1/3$, but note that our results are reported in dimensionless units and are hence valid on any scale.

\subsection*{Experiments}

In the experiments shown in Fig.~\ref{fig:phase_diagram}, we tangentially attached straight steel pipes to rigid polystyrene spheres and manually fed polyurethane wires from both sides at equal speed through the pipes into the spheres. Starting from an initially preset loop, the wire then developed into the spiral or classical morphology depending on friction, which we controlled with a silicone lubricant. For the morphologies in weak confinement, we used stiffer polycaprolactam wires and customary stretchable balloons made of natural rubber into which the wires were tangentially pushed by hand.

\subsection*{Acknowledgements}

Financial support from ETH Zurich by ETHIIRA Grant No.~ETH-03 10-3 as well as from the European Research Council (ERC) Advanced Grant No.~319968-FlowCCS is gratefully acknowledged. We thank N.~Stoop for valuable discussions.

\end{multicols}

\normalsize

\subsection*{Supplementary Note 1: Theoretical framework}
\label{sec:suppl1}

In a Kirchhoff kinematic Cosserat description, a thin filament of length $L$ can be represented by its parametric centreline, $\Gamma: s\in [0,L]\mapsto\vec{p}(s)\in\mathbb{E}^3$, and an orthonormal director frame $\vec{d}_i, i=1,2,3$, specifying the cross-sectional orientation along $\Gamma$, with $\vec{d}_3=\partial_s\vec{p}/|\partial_s\vec{p}|$ \cite{CC09}. The Darboux vector $\vec{k}$ is uniquely defined by $\partial_s\vec{d}_i=\vec{k}\times\vec{d}_i$. The elastic strain energy $U_{\mathrm f}$ of the filament, whose isotropic homogeneous elastic material is characterized by Young's modulus $E_{\mathrm f}$ and Poisson's ratio $\nu_{\mathrm f}$, can be compactly expressed by
\begin{equation}
U_{\mathrm f} = \frac{1}{2} \int_\Gamma\mathrm ds\,\left[E_{\mathrm f}I_{\mathrm f}\left(\vartheta_1^2+\vartheta_2^2\right) + G_{\mathrm f}J_{\mathrm f}\vartheta_3^2 + E_{\mathrm f}A_{\mathrm f}\eta^2\right],
\label{eq:filament}
\end{equation}
where $\vartheta_i=k_i-\overline{k}_i$, $k_i=\vec{k}\cdot\vec{d}_i$, $\eta=u-\overline{u}$, $u=\partial_s\vec{p}\cdot\vec{d}_3$. Barred symbols refer to an undeformed stress-free reference configuration, with respect to which the deformed configuration (bare symbols) is described. For a circular cross section with radius $r$, the cross section area is $A_{\mathrm f}=\pi r^2$, the area moment of inertia $I_{\mathrm f}=\pi r^4/4$ and the polar moment of inertia $J_{\mathrm f}=\pi r^4/2$. In linear elasticity, the shear modulus reads $G_{\mathrm f}=E_{\mathrm f}/2(1+\nu_{\mathrm f})$. The respective terms in equation (\ref{eq:filament}) account for bending in two directions ($U_{\mathrm b}$), torsion ($U_{\mathrm t}$), and axial tension or compression.

In a total Lagrangian description of a Kirchhoff\textendash{}Love thin sheet with small thickness $h$, whose isotropic homogeneous elastic material is characterized by Young's modulus $E_{\mathrm s}$ and Poisson's ratio $\nu_{\mathrm s}$, and which obeys the geometrically nonlinear St.~Venant-Kirchhoff law, the total elastic energy reads \cite{K66}
\begin{equation}
U_{\mathrm s} = \frac{1}{2} \int_{\overline{\Omega}}\mathrm d\overline{\Omega}\,\left[ \alpha_{\alpha\beta}C^{\alpha\beta\gamma\delta}\alpha_{\gamma\delta} + \frac{h^2}{12} \beta_{\alpha\beta}C^{\alpha\beta\gamma\delta}\beta_{\gamma\delta} \right],
\label{eq:shell}
\end{equation}
where $C^{\alpha\beta\gamma\delta}$ are the components of the elastic tensor in curvilinear coordinates integrated over the constant sheet thickness $h$, given by
\begin{equation}
C^{\alpha\beta\gamma\delta} = K\left(\nu_{\mathrm s}\overline{a}^{\alpha\beta}\overline{a}^{\gamma\delta}+\frac{1-\nu_{\mathrm s}}{2}\left[\overline{a}^{\alpha\gamma}\overline{a}^{\beta\delta}+\overline{a}^{\alpha\delta}\overline{a}^{\beta\gamma}\right]\right).
\end{equation}
$K = E_{\mathrm s}h/(1-\nu_{\mathrm s}^2)$ is the stretching modulus. $\overline{\Omega}$ denotes the sheet's parametric middle surface in a stress-free reference state, again denoted by barred quantities. Einstein summation over the Greek indices $\alpha,\beta,\gamma,\delta=1,2$ is implied, with subscripts and superscripts denoting covariant and contravariant components, respectively. The residual membrane strains $\alpha_{\alpha \beta} = (a_{\alpha\beta} - \overline{a}_{\alpha\beta})/2$ are the covariant components of the difference between the first fundamental forms of the middle surface in deformed ($a$) and reference ($\overline{a}$) configuration, and the residual bending strains $\beta_{\alpha\beta} = \overline{b}_{\alpha\beta} - b_{\alpha\beta}$ derive analogously from the second fundamental forms $b$ and $\overline{b}$, respectively.

\subsection*{Supplementary Note 2: Analytical approximation for the spiral morphology}
\label{sec:suppl2}

\begin{figure*}[h]
	\centering
	\includegraphics[height=4cm]{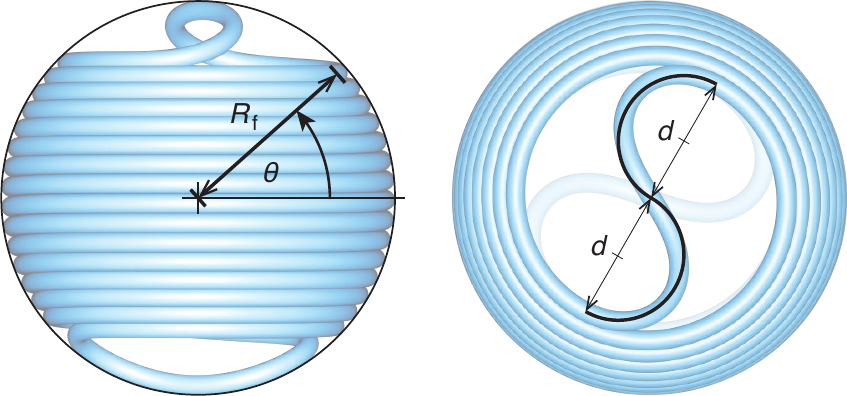}
	\caption{\textbf{Geometrical approximation of the spiral morphology.} The filament can be divided into three parts: a coil with inclination $\theta$ (left) and two S-shaped curves (right).}
	\label{fig:spiral}
\end{figure*}

Frictionless rigid spherical confinements give rise to chiral surface-covering filament packings consisting of a dense coil and two spiral S-curves located at the poles (Fig.~\ref{fig:spiral}). This high degree of order grants access to an approximate closed-form solution of the predominant elastic contribution, the bending energy of the confined filament, beyond the point of first contact ($l>l^{*}$). Even for moderate system sizes $\sigma$, we can assume inextensibility of the filament and neglect the pitch of the helical coil. Its length $L^{\mathrm{coil}}$ can thus be estimated by summing up rings over an equatorial band on the surface of a sphere with radius $R_{\mathrm f}$. The angular width of the band is $2\theta$, where $\theta$ is the inclination of the coil as depicted in the figure, yielding
\begin{equation}
L^{\mathrm{coil}} \approx \frac{1}{2r}\int_{-\theta}^\theta\mathrm d\theta'\int_0^{2\pi}\mathrm d\varphi\,R_{\mathrm f}^2\cos(\theta') = 2\pi\frac{R_{\mathrm f}^2}{r}\sin(\theta).
\end{equation}
Analogously, the bending energy of the coil reads
\begin{equation}
U_{\mathrm{b}}^{\mathrm{coil}} \approx \frac{1}{2r}\int_{-\theta}^\theta\mathrm d\theta'\int_0^{2\pi}\mathrm d\varphi\,\frac{E_{\mathrm f}I_{\mathrm f}R_{\mathrm f}}{2R_{\mathrm f}\cos(\theta')} = \pi\frac{E_{\mathrm f}I_{\mathrm f}}{r}\log\left[1+\frac{2}{\cot(\theta/2)-1}\right].
\end{equation}
A rough estimate for the contribution from the two central spiral patterns is given by four semicircles with a diameter of $d=R_{\mathrm f}(\pi/2-\theta)$. (An alternative estimate can be found in ref.~\cite{BABCCET06}.) Adding these contributions up, one finds the total filament length $L \approx L^{\mathrm{coil}} + 2\pi d$ and its total bending energy
\begin{equation}
U_{\mathrm b} \approx U_{\mathrm{b}}^{\mathrm{coil}} + 4\pi\frac{E_{\mathrm f}I_{\mathrm f}}{d}.
\label{eq:ebend}
\end{equation}
To obtain the bending energy as a function of the normalized filament length $l$, the coil inclination $\theta$ can be eliminated from equation (\ref{eq:ebend}) by numerically solving
\begin{equation}
l = \frac{L}{2\pi R_{\mathrm f}} \approx \frac{R_{\mathrm f}}{r}\sin(\theta)+\pi/2-\theta
\end{equation}
for $\theta$. Finally, the number of windings in the coil is $\theta R_{\mathrm f}/r$.

\subsection*{Supplementary Note 3: Folding in weak confinements}
\label{sec:suppl3}

\begin{figure*}[ht]
	\centering
	\includegraphics[width=\textwidth]{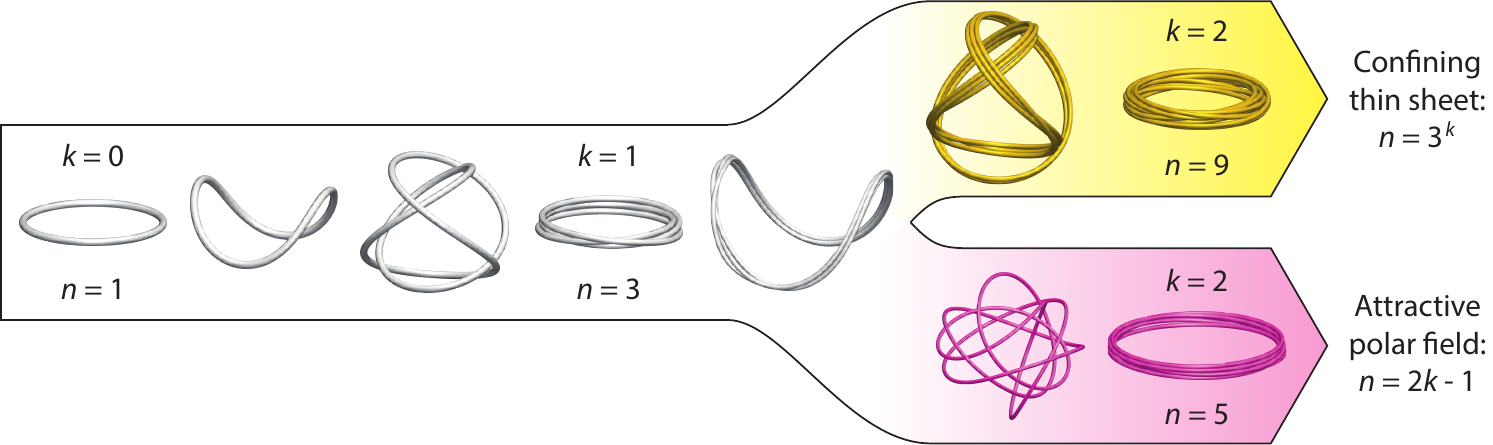}
	\caption{\textbf{Folding of a filament loop in weak confinements.} The folding behaviour of growing filament loops strongly depends on the type of confinement. The number of windings $n$ scales exponentially with the number of folds $k$ in thin elastic sheets, and linearly in polar attractive force fields.}
	\label{fig:folded_vs_polar_field}
\end{figure*}

The exponential scaling of the winding number $n=(2m-1)^k$ in the folded morphology is a particular feature of the type of weak confinement considered here. By replacing the confining thin sheet by an attractive polar force field with potential $\Phi(r)\sim r^{p}$, $p\geq 1$, where $r$ is the distance from the centre, we found a cascade of folds with winding number $n=2k-1$ for the stable buckling mode $m=2$ (Fig.~\ref{fig:folded_vs_polar_field}) similar to phantom rings confined to the surface of a sphere \cite{GV12}. This dramatic difference in scaling stems from the anisotropic spatial confinement exerted by a flexible sheet: In a polar field, the single strands in the buckling rope have room to separate and disentangle during the transition to a higher winding number, resulting in the low energy modes $n=2k-1$. This is prevented by the interaction with an elastic thin sheet, which wraps around the folded strands, enforcing the exponential law by keeping the bundle together.

Spatial confinement is not the only way of obtaining the saddle shape shown in Fig.~\ref{fig:folded_vs_polar_field} from a buckled single-stranded ring. The necessary excess curvature can also be imposed by creasing an annulus or by attaching two open ends of a ring at an excess angle \cite{MCPRJ12}. Without the flexible confinement exerted by an enclosing thin sheet, however, the cascade of self-similar folds that we observe in the folded phase has not been reported before. An unconstrained overcurved ring or annulus will just wind $n=2k-1$ times.

\begin{multicols}{2}

\end{multicols}

\end{document}